\documentclass[12pt]{iopart}
\expandafter\let\csname equation*\endcsname=\relax
\expandafter\let\csname endequation*\endcsname=\relax
\usepackage{graphicx}
\usepackage{amsmath}
\usepackage{braket}

\begin{document}

\title[IOP Publishing journals]{Synthetically enhanced sensitivity using higher-order exceptional point and coherent perfect absorption}

\author
{Yao-Dong Hu,$^{1}$ Yi-Pu Wang,$^{1\ast}$ Rui-Chang Shen,$^{1}$ Zi-Qi Wang,$^{1}$ Wei-Jiang Wu,$^{1}$ J. Q. You$^{1,2\dagger}$}
\address{$^{1}$ Zhejiang Province Key Laboratory of Quantum Technology and Device, School of Physics, and State Key Laboratory for Extreme Photonics and Instrumentation, Zhejiang University, Hangzhou 310027, China}
\address{$^{2}$ College of Optical Science and Engineering, Zhejiang University, Hangzhou 310027, China}

\ead{yipuwang@zju.edu.cn and jqyou@zju.edu.cn}
\vspace{10pt}
\begin{indented}
\item[]November 2023
\end{indented}

\begin{abstract}
Sensors play a crucial role in advanced apparatuses and it is persistently pursued to improve their sensitivities. Recently, the singularity of a non-Hermitian system, known as the exceptional point (EP), has drawn much attention for this goal. Response of the eigenfrequency shift to a perturbation $\epsilon$ follows the $\epsilon^{1/n}$-dependence at an $n$th-order EP, leading to significantly enhanced sensitivity via a high-order EP. However, due to the requirement of increasingly complicated systems, great difficulties will occur along the path of increasing the EP order to enhance the sensitivity. Here we report that by utilizing the spectral anomaly of the coherent perfect absorption (CPA), the sensitivity at a third-order EP can be further enhanced owing to the cooperative effects of both CPA and EP. We realize this synthetically enhanced sensor using a pseudo-Hermitian cavity magnonic system composed of two yttrium iron garnet spheres and a microwave cavity. The detectable minimum change of the magnetic field reaches $4.2\times 10^{-21}$~T. It opens a new avenue to design novel sensors using hybrid non-Hermitian quantum systems.
\end{abstract}

%
%
%
%
%

\section{Introduction}
\noindent Non-Hermitian systems are ubiquitous in nature, owing to the flow of energy, particles, and information to and from the external degrees of freedom that lie outside the Hilbert space of the considered system~\cite{Bender2007}. They exhibit characteristics distinct from the Hermitian counterparts~\cite{Rotter2009,Shen2018,Leykam2017,Bergholtz2021}. A critical advancement of the non-Hermitian physics is related to the singularity of the system, known as the exceptional point (EP)~\cite{Heiss2004,Heiss2012}, where eigenvalues become degenerate and the corresponding eigenvectors coalesce. This singularity brings new possibilities for intriguing applications, including the optical isolation~\cite{Chrisrian2010,Chang2014,Peng2014}, band merging~\cite{Makris2008,Zhen2015}, loss-induced transparency~\cite{Guo2009}, asymmetric mode switching~\cite{Doppler2016}, single-mode lasers~\cite{Brandstetter2014,Lin2016,Feng2014,Hodaei2014}, and topological chirality~\cite{Dembowski2001,Xu2017,Zhou2018,Xu2016,Hassan2017}.
Novel properties at or around the EP have been achieved in a vast array of platforms, such as microwave~\cite{Dembowski2001,Peng2014}, photonic~\cite{Makris2008,Zhen2015,Guo2009,Doppler2016,Brandstetter2014,Lin2016,Xu2017,Zhou2018,Ozdemir2019,Miri2019,Parto2020}, magnetic~\cite{Zhang2017,Liu2019}, acoustic~\cite{Ding2016}, mechanical~\cite{Xu2016}, and optomechanical systems~\cite{PhysRevLett.113.053604,PhysRevLett.114.253601}.

Recently, EP-based sensors have attracted considerable interest due to the enhanced response to the external perturbation, providing an efficient way to boost detection sensitivity~\cite{Wiersig2014,Hodaei2017,Chen2017,
Zhong2019,Lai2019,Wiersig2020,Tang2020,Budich2020}. The eigenfrequency splitting at the $n$th-order EP (i.e., EP$n$) follows a $\epsilon^{1/n}$-dependence (where $\epsilon$ denotes the perturbation amplitude), which indicates that a higher-order EP is beneficial to higher sensitivity. Indeed, successful demonstrations of the enhanced sensitivity have been realized at the EP3~\cite{Hodaei2017,Chen2017}, but great difficulties will occur along the path of further increasing the EP order to enhance the sensitivity. This is because the achievement of higher-order EPs requires more precise parameter controls. Thus, to further increase the sensitivity of EP-based sensors, it is necessary to contemplate new mechanisms.
In this work, we achieve the EP3 of a pseudo-Hermitian Hamiltonian~\cite{Mostafazadeh2002,Mostafazadeh2002-1}, which is implemented using a cavity magnonic system via precisely adjusting multi-parameters~\cite{Zhang2017,ziqi2022}. The coherent perfect absorption (CPA)~\cite{Chong2010,PhysRevLett.106.093902,Sun2014,Baranov2017,Soleymani2022,Zhang2023} is also used to further enhance the sensitivity at the EP3. The resulting sensitivity is remarkably enhanced owing to the {\it cooperative} effects of both CPA and EP3. A formulated explanation of this synthetic sensitivity can be characterized by the detectable minimum change of the magnetic field,
\begin{equation}\label{ale'}
		\delta B_{\rm min} = 2\pi\frac{\delta A}{\gamma_e G_{\rm CPA} G_{\rm EP3}},
\end{equation}
where $\gamma_e$ is the the gyromagnetic ratio of the ferrimagnetic material that we use ($\gamma_e =2\pi*28$~GHz/T), and $\delta A$ represents the smallest spectrum change that can be resolved in the experiment (e.g., $1\times 10^{-13}$~dB). $G_{\rm CPA}$ and $G_{\rm EP3}$ are sensitivity factors associated with the CPA and EP3, respectively. It is clear that the sensitivity induced by these two mechanisms is superimposed in a {\it product} form, thus yielding a vast enhancement. Our work paves a novel way to greatly enhance the sensitivity using both CPA and EP in a single non-Hermitian system.

\section{Pseudo-Hermitian cavity magnonic system}
\subsection{Experimental setup}
\begin{figure}
	\centering
	\includegraphics[width=1\textwidth]{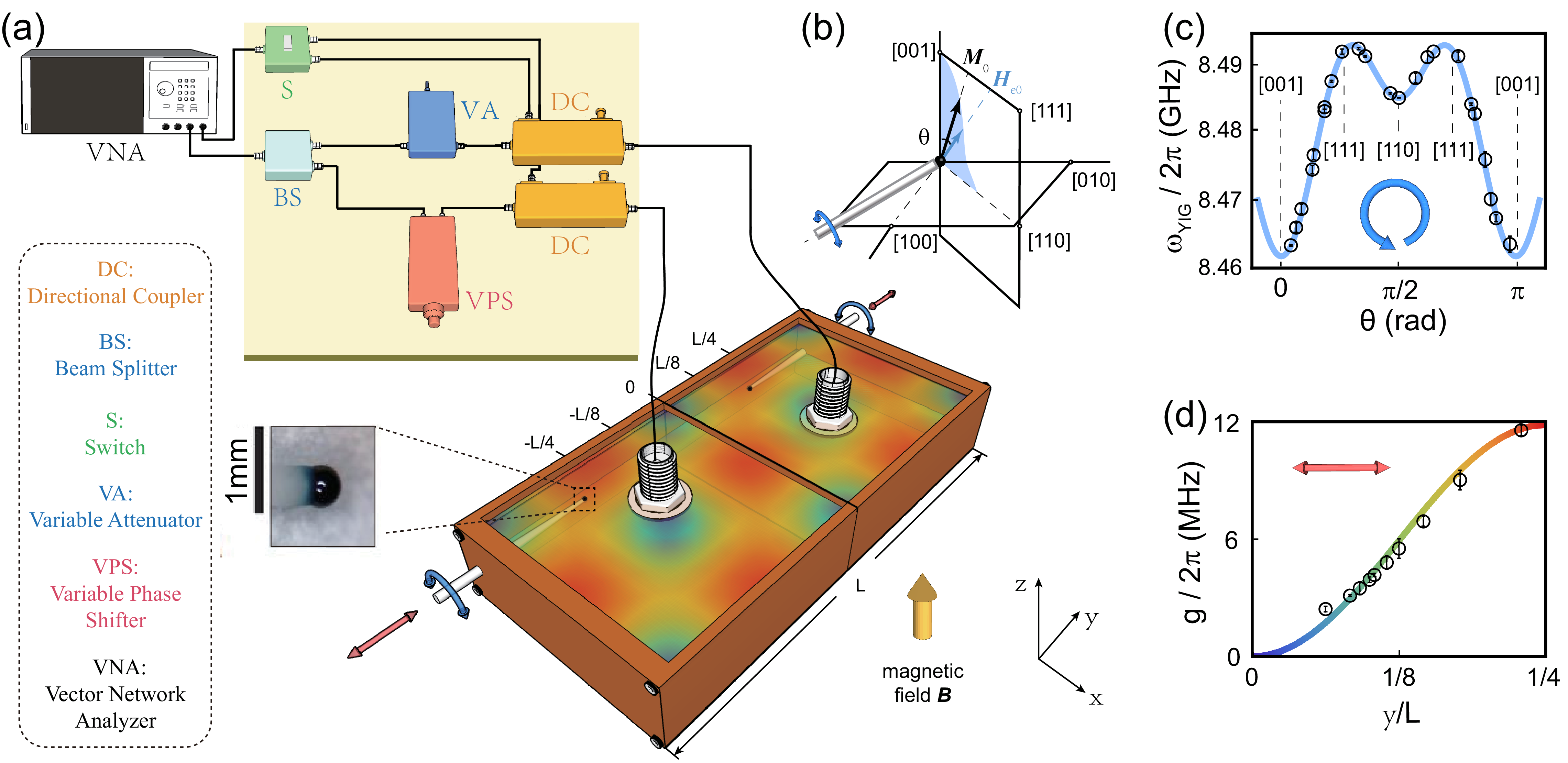}
	\caption{Schematic diagrams of the experimental setup and parameter adjustments. (a) The microwave signal generated by a vector network analyzer (VNA) is divided into two signals by a beam splitter (BS). These two signals are subsequently tuned by a variable attenuator (VA) and a variable phase shifter (VPS), respectively, to match the CPA condition. Then, the signals are loaded into the cavity via two directional couplers (DC), by which the output signals are sent back to the VNA. Two 0.5~mm-diameter YIG spheres are placed in the cavity. The magnetic-field distribution of the cavity mode $\rm{TE}_{102}$ is shown as a colour diagram. (b) The magnetocrystalline anisotropy field is employed to tune the frequency of the magnon mode. (c) The frequency of the magnon mode (circles) is measured by rotating the YIG sphere. (d) The coupling strength between the magnon mode and the cavity mode (circles) is tuned by moving the YIG sphere from 0 to $L$/4, where $L$ is the length of the cavity. The curves in (c) and (d) are theoretical results.}\label{f1}
\end{figure}

The cavity magnonic system consists of a three-dimensional (3D) rectangular cavity ($50\times25\times7.5~\rm{mm}^3$) and two 0.5~mm-diameter yttrium  iron garnet (YIG) spheres [figure~\ref{f1}(a)]. The two ports of the cavity are connected to a vector network analyzer (VNA) for both loading microwave signals and implementing measurement. The power ratio and phase difference of the two signals are controlled by a variable attenuator (VA) and a variable phase shifter (VPS), respectively. Here, the CPA is realized by precisely adjusting the VA and VPS to achieve zero output signals from the cavity. The dissipation rates of the cavity ports are tuned by changing the lengths of antenna pins inserted into the cavity~(\ref{app1}).
Each YIG sphere is glued on a plastic rod to adjust its position and crystal axis orientation. After applying a bias magnetic field $\textbf{\textit{B}}$ in the $z$-direction, magnon modes are supported in the sphere. Here we focus on the Kittel mode, i.e., the uniform precession mode of the spin ensemble. By moving the YIG sphere along the edge of the cavity ($y$-direction), the coupling strength between the magnon and cavity modes can be precisely controlled. Meanwhile, rotating the rod can change the frequency of the magnon mode owing to the magnetocrystalline anisotropy~\cite{ziqi2022}. Specifically, rotating the rod from 0$^\circ$ to 180$^\circ$ changes the magnon frequency in a range of 0.3~GHz [figures~\ref{f1}(b) and~\ref{f1}(c)]. When the position of the YIG sphere is tuned from 0 to $\pm L/4$, the coupling strength between the magnon and cavity modes can be tuned from 0 to 12~MHz [figure~\ref{f1}(d)]. With these control parameters, we can simultaneously realize the CPA and EP3 in the system.

\subsection{Pseudo-Hermitian system with a third-order exceptional point}
The non-Hermitian Hamiltonian of the system with CPA can be written, in the rotating frame associated with the cavity frequency (details in \ref{app2}):
\begin{equation}
		H/\hbar=
		\left(
		\begin{matrix}
			i\kappa_c & g_1 & g_2\\
			g_1 & \Delta_1-i\gamma_1 & 0\\
			g_2 & 0 & \Delta_2-i\gamma_2\\
		\end{matrix}\right),
		\label{Ham1}
	\end{equation}
where $\kappa_{c} \equiv \kappa_1 + \kappa_2 - \kappa_{\rm int}>0$ is the effective gain of the cavity mode, $g_1$ ($g_2$) is the coupling strength between the cavity mode and the magnon mode in the YIG sphere 1 (2), which has decay rate $\gamma_1$ ($\gamma_2$), and $\Delta_1~(\Delta_2) = \omega_1~(\omega_2)-\omega_{\rm{c}}$ is the frequency detuning between the cavity mode and the corresponding magnon mode.

\begin{figure}
	\centering
	\includegraphics[width=0.6\textwidth]{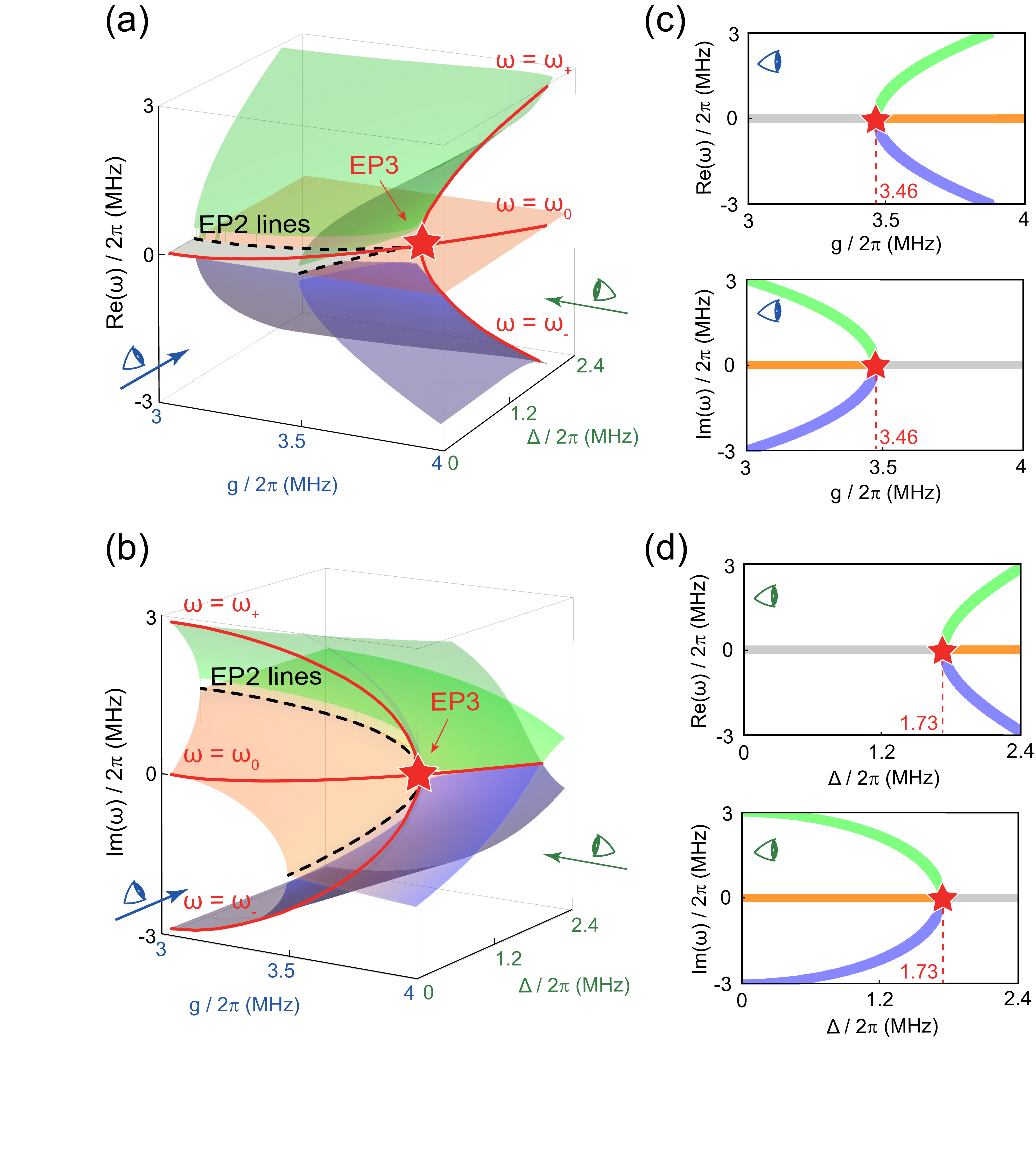}
	\caption{Eigenvalues of the three-mode non-Hermitian system. (a) and (b) Real and imaginary parts of the three eigenvalues versus the coupling strength $g/2\pi$ and frequency detuning $\Delta/2\pi$, as illustrated by the green, orange and purple surfaces, respectively. The EP2 lines are indicated as the black dashed curves, and the intersection point of the two curves is the EP3 (red star). (c) Real and imaginary parts of the eigenvalues versus $g/2\pi$, corresponding to the side views of (a) and (b) by the ``blue" eye. The value of $g/2\pi$ at EP3 is $g_{\rm EP3}/2\pi = 3.46$~MHz. (d) Real and imaginary parts of the eigenvalues versus $\Delta/2\pi$, corresponding to the side views of (a) and (b) by the ``green" eye. The value of $\Delta/2\pi$ at EP3 is $\Delta_{\rm EP3}/2\pi = 1.73$~MHz. (c) and (d) give the three eigenvalues of the pseudo-Hermitian Hamiltonian.}\label{f2}
\end{figure}

In the symmetric case, we have two identical YIG spheres ($\gamma_1=\gamma_2=\gamma$) and the same coupling strengths $g_1=g_2=g$. Also, the effective gain of the cavity is tuned to $\kappa_{c}=2\gamma$, and $\Delta_1=-\Delta_2\equiv\Delta$. Then, the Hamiltonian reduces to
	\begin{equation}\label{Ham2}
		H/\hbar =
		\left(
		\begin{matrix}
			2i\gamma & g & g\\
			g & \Delta-i\gamma & 0\\
			g & 0 & -\Delta-i\gamma\\
		\end{matrix}\right).
	\end{equation}
The eigenvalues of this Hamiltonian are determined by~${\rm Det}\left( H - \Omega I \right) = 0$, which gives the characteristic polynomial,
\begin{equation}\label{poly}
\Omega^{3} + c_1\Omega + c_0= 0,
\end{equation}
with $c_0 = 2 i\gamma(\Delta^2+\gamma^2-g^2)$, and $c_1 = 3\gamma^2-2g^2-\Delta^2$. Here $\Omega\equiv\omega-\omega_c$ is the biased eigenvalue of the system and $I$ is the identity matrix.
According to the Cardano formula, the eigenvalues of the system can be explicitly written as
\begin{align}\label{w0}
\Omega^{(0)} &= \sqrt[3]{\frac{c_0}{2}+\sqrt{\frac{c_0^2}{4}+\frac{c_1^3}{27}}}+\sqrt[3]{\frac{c_0}{2}-\sqrt{\frac{c_0^2}{4}+\frac{c_1^3}{27}}},\\
\Omega^{(1)} &= \sqrt[3]{\frac{c_0}{2}+\sqrt{\frac{c_0^2}{4}+\frac{c_1^3}{27}}}
\left(-\frac{1}{2}+i\frac{\sqrt{3}}{2}\right)-\sqrt[3]{\frac{c_0}{2}-\sqrt{\frac{c_0^2}{4}+\frac{c_1^3}{27}}}
\left(-\frac{1}{2}+i\frac{\sqrt{3}}{2}\right)^2,\nonumber\\
\Omega^{(2)} &= \sqrt[3]{\frac{c_0}{2}+\sqrt{\frac{c_0^2}{4}+\frac{c_1^3}{27}}}
\left(-\frac{1}{2}+i\frac{\sqrt{3}}{2}\right)^2-\sqrt[3]{\frac{c_0}{2}-\sqrt{\frac{c_0^2}{4}+\frac{c_1^3}{27}}}
\left(-\frac{1}{2}+i\frac{\sqrt{3}}{2}\right).\nonumber
\end{align}
In figures~\ref{f2}(a) and~\ref{f2}(b), we plot these three eigenvalues versus both $g$ and $\Delta$.

(i)~When $g=\sqrt{\Delta^2+\gamma^2}$, i.e., $c_0=0$, the three eigenvalues in equation~(\ref{w0}) are reduced to
\begin{equation}\label{w1}
\Omega^{(0)} = 0,~~~~~
\Omega^{(\pm)} =\pm \sqrt{3g^{2} - 4\gamma^{2}},
\end{equation}
which are drawn as the red curves in figures~\ref{f2}(a) and~\ref{f2}(b). Note that the non-Hermitian Hamiltonian (\ref{Ham2}) with $g=\sqrt{\Delta^2+\gamma^2}$ does not have the parity-time (PT) symmetry, but can still have three real eigenvalues when $g>g_{\rm EP3}\equiv 2\gamma/\sqrt{3}$. This is because the non-Hermitian Hamiltonian becomes {\it pseudo-Hermitian} in this case~\cite{Mostafazadeh2002,Mostafazadeh2002-1,Kawabata2019,Rivero2020}.
When $g<g_{\rm EP3}$, $\Omega^{(\pm)}$ become complex, i.e., one of the three eigenvalues is real ($\Omega^{(0)} = 0$) and the other two, $\Omega^{(\pm)}$, are a complex-conjugate pair. In particular, when $g=g_{\rm EP3}\equiv 2\gamma/\sqrt{3}$, there is $\Delta=\Delta_{\rm EP3}\equiv\gamma/\sqrt{3}$ owing to the relation $g=\sqrt{\Delta^2+\gamma^2}$. Now, the three eigenvalues coalesce to $\Omega^{(\pm)}=\Omega^{(0)}=0$. In figures~\ref{f2}(c) and~\ref{f2}(d), these three eigenvalues of the pseudo-Hermitian Hamiltonian are shown versus $g$ and $\Delta$, respectively.

Corresponding to the three eigenvalues in equation~(\ref{w1}), the pseudo-Hermitian~Hamiltonian, i.e., the non-Hermitian Hamiltonian (\ref{Ham2}) with $g=\sqrt{\Delta^2+\gamma^2}$, have the following eigenvectors:
\begin{align}\label{eignvector}
&\ket{\Phi_0} =
\left(
\begin{array}{ccc}
1\\
\frac{-\sqrt{g^2-\gamma^2}-i\gamma}{g}\\
\frac{\sqrt{g^2-\gamma^2}-i\gamma}{g}
\end{array}
\right), \nonumber\\
&\ket{\Phi_+} =
\left(
\begin{array}{ccc}
1\\
\frac{g}{\sqrt{3g^2-4\gamma^2}-\sqrt{g^2-\gamma^2}+i\gamma}\\
\frac{g}{\sqrt{3g^2-4\gamma^2}+\sqrt{g^2-\gamma^2}+i\gamma}
\end{array}
\right), \nonumber\\
&\ket{\Phi_-} =
\left(
\begin{array}{ccc}
1\\
\frac{g}{-\sqrt{3g^2-4\gamma^2}-\sqrt{g^2-\gamma^2}+i\gamma}\\
\frac{g}{-\sqrt{3g^2-4\gamma^2}+\sqrt{g^2-\gamma^2}+i\gamma}
\end{array}
\right).
\end{align}
At $g=g_{\rm EP3}\equiv 2\gamma/\sqrt{3}$, these three eigenvectors coalesce to
\begin{equation}
\ket{\Phi_{\pm}}=\ket{\Phi_0} =
\left(
\begin{array}{ccc}
1\\
\frac{-1-i\sqrt{3}}{2}\\
\frac{1-i\sqrt{3}}{2}
\end{array}
\right),
\end{equation}
revealing that $g_{\rm EP3}\equiv 2\gamma/\sqrt{3}$ is indeed the third-order exceptional point (EP3) of the pseudo-Hermitian Hamiltonian.

(ii)~The three eigenvalues in equation~(\ref{w0}) have EP2 lines when $27c_0^2+4c_1^3=0$. In this case, the three eigenvalues reduce to
\begin{align}\label{w2}
&\Omega^{(0)} = 2\sqrt[3]{\frac{c_0}{2}},\nonumber\\
&\Omega^{(1)} = i\sqrt{3}\sqrt[3]{\frac{c_0}{2}},\nonumber\\
&\Omega^{(2)}= -i\sqrt{3}\sqrt[3]{\frac{c_0}{2}},
\end{align}
Furthermore, when $c_0=0$, it follows from $27c_0^2+4c_1^3=0$ that $c_1=0$ as well, which corresponds to the critical parameters $g=g_{\rm EP3}\equiv 2\gamma/\sqrt{3}$, and $\Delta=\Delta_{\rm EP3}\equiv\gamma/\sqrt{3}$. Now, the three eigenvalues in equation~(\ref{w2}) coalesce to $\Omega^{(0)}=\Omega^{(1)}=\Omega^{(2)}=0$. This corresponds to the case where the EP2 lines insect at the EP3, as shown in figures~\ref{f2}(a) and~\ref{f2}(b).

\section{Results and discussions}

\subsection{Output spectra at and away from EP3}

\begin{figure}
	\centering
	\includegraphics[width=1.05\textwidth]{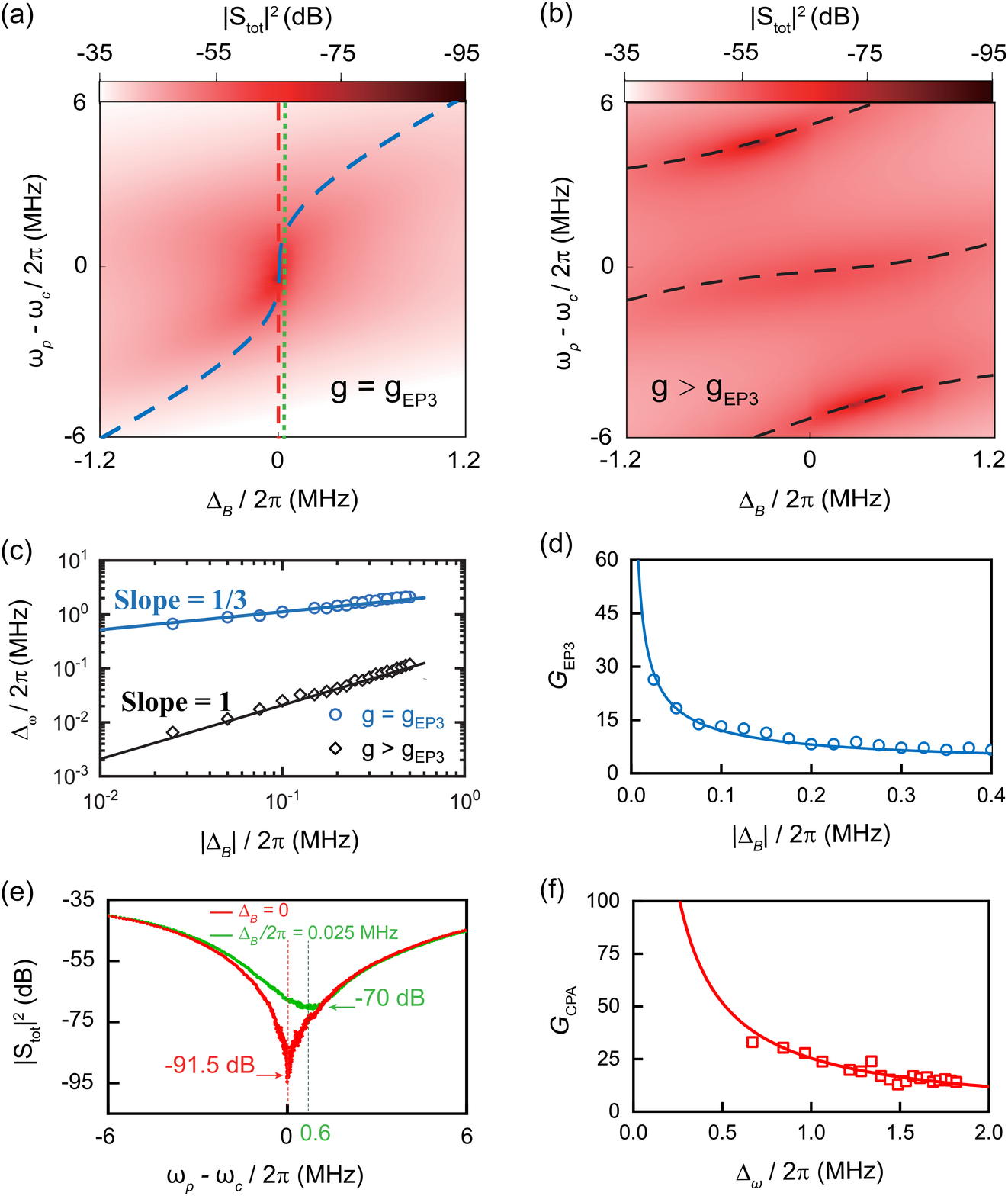}
	\caption{Output spectra and sensitivity factors. (a) Total output spectrum at the EP3 ($g=g_{\rm EP3}=2\pi\times 3.46$~MHz) versus the perturbation $\Delta_B/2\pi$, where the dashed and dotted vertical lines correspond to $\Delta_B/2\pi=0$ and $0.025$~MHz, respectively. (b) Total output spectrum away from the EP3 ($g=2\pi\times 4.59~{\rm MHz}>g_{\rm EP3}$) versus the perturbation $\Delta_B/2\pi$. Dashed curves in (a) and (b) are eigenfrequency shifts $\Delta_{\omega}=\Omega'-\Omega$ obtained by diagonalizing the Hamiltonian, i.e.~equation (\ref{H'}). (c) Log-log plot of the frequency shift $\Delta_\omega/2\pi$ of the central eigenvalue versus $|\Delta_B|/2\pi$ at $g/2\pi=3.46$ and $4.59$~MHz, respectively. The slopes 1/3 and 1 are fitted using equations~(\ref{pertb3}) and (\ref{pertb4}), respectively. (d) Sensitivity factor at the EP3 versus $|\Delta_B|/2\pi$. (e) Total output spectra extracted at $\Delta/2\pi=0$ and $0.025$~MHz, corresponding to the two vertical lines in (a). The spectral dip at $\Delta_B=0$ has the minimum value $-91.5$~dB. (f) CPA-related sensitivity factor $G_{\rm CPA}$ versus the eigenfrequency shift $\Delta_\omega/2\pi$.}\label{f3}
\end{figure}

In the experiment, we utilize a small variation of the external magnetic field as the perturbation, which results in a frequency change $\Delta_B$ for the Kittel mode of each YIG sphere. The perturbed pseudo-Hermitian Hamiltonian has been shown in equation~(\ref{H'}), where the system has the eigenvalue shift $\Delta_\omega\equiv\Omega'-\Omega$ under the perturbation $\Delta_B$. In figures~\ref{f3}(a) and~\ref{f3}(b), we measure the output spectra to demonstrate the response of the pseudo-Hermitian system to $\Delta_B$ {\it at} and {\it away from} the EP3. The darkest red areas correspond to the system with (or nearly with) CPA under zero (or small) perturbation $\Delta_B$. It is clear that each of the three eigenvalue has a frequency shift in the case of $g>g_{\rm EP3}$ when applying the perturbation $\Delta_B$ [figure~\ref{f3}(b)], while these frequency shifts become the same at the EP3 ($g=g_{\rm EP3}$) because all eigenvalues coalesce there [figure~\ref{f3}(a)]. Hereafter, we focus on the central eigenvalue $\Omega=0$ of the pseudo-Hermitian Hamiltonian, so $\Omega'=\Delta_\omega$.

\subsection{Perturbation-induced eigenfrequency shift}

The perturbation of the external magnetic field can induce frequency change of the Kittel mode in each YIG sphere. In our experiment, we use two identical YIG spheres, so these frequency changes can be nearly identical. By considering the magnetic-field perturbation, the pseudo-Hermitian Hamiltonian, i.e., the Hamiltonian (\ref{Ham2}) with $\Delta=\sqrt{g^2-\gamma^2}$, can be converted to
\begin{equation}\label{H'}
H'/\hbar = \begin{pmatrix}
2i\gamma & g & g \\
g & \sqrt{g^2-\gamma^2}+ \Delta_B - i\gamma & 0 \\
g & 0 & -\sqrt{g^2-\gamma^2}+ \Delta_B - i\gamma \\
\end{pmatrix},
\end{equation}
where $\Delta_{B}$ is the frequency change of the Kittel mode induced by the magnetic-field perturbation. From the eigenvalue equation ${\rm Det}\left(H' - \Omega'I \right) = 0$, we can obtain the relation between the eigenvalue $\Omega'(\Delta_B) \equiv\omega'(\Delta_B)-\omega_c$ and the frequency change $\Delta_B$, which is
\begin{equation}
\frac{2\Delta_B - \Omega'}{g^2} + \frac{4\left( g^{2} - \gamma^{2} \right)\left( \Omega' - \Delta_{B} \right)}{\left\lbrack {\left( \Omega' - \Delta_{B} \right)^{2} + g^{2}} \right\rbrack^{2} - 4\left( g^{2} - \gamma^{2} \right)\left( \Omega' - \Delta_{B} \right)^{2}} = 0,
\end{equation}
where $\Omega'(\Delta_B) \equiv\omega'(\Delta_B)-\omega_c$ reduces to the eigenvalue $\Omega \equiv\omega-\omega_c$ of the pseudo-Hermitian Hamiltonian when $\Delta_B=0$.

For a sufficiently small perturbation $\Delta_{B}$, we can ignore the higher-order terms $O\left( \Delta_{B}^{2} \right)$ and write the relation between $\Delta_{B}$ and $\Omega'$ as
\begin{equation}\label{pertb}
\Delta_{B} = \frac{\Omega'^{3}\left( \Omega'^{2} + g^{2} \right) - \left( 3g^{2}-4\gamma^{2} \right)\left( \Omega'^{2} + g^{2} \right)\Omega'}{g^{4} + 2\Omega'^{2}\left( \Omega'^{2} + 2g^{2} \right) - \left( 3g^{2}-4\gamma^{2} \right)\left( 4\Omega'^{2} + g^{2} \right)}.
\end{equation}
Based on equation~(\ref{pertb}), we can study the response of the eigenfrequency to the perturbation.

We define the perturbation-induced eigenfrequency shift as
\begin{equation}\label{del&o&o'}
\Delta_{\omega} \equiv \omega'\left( \Delta_{B} \right) - \omega = \Omega'\left( \Delta_{B} \right) - \Omega.
\end{equation}
Here we focus on the {\it central} branch of the eigenvalues of the pseudo-Hermitian Hamiltonian, i.e., $\Omega=\Omega^{(0)}=0$ [cf. equation~(\ref{w1})].

(i)~At the EP3, i.e., $g=g_{\rm EP3}\equiv 2\gamma/\sqrt{3}$, equation~(\ref{pertb}) reduces to
\begin{equation}\label{pertb1}
\Delta_{B} = \frac{\Omega'^{3}\left( \Omega'^{2} + g_{\rm EP3}^{2} \right) }{g_{\rm EP3}^{4} + 2\Omega'^{2}\left( \Omega'^{2} + 2g_{\rm EP3}^{2} \right)}.
\end{equation}
For a small perturbation, we can assume that $\Omega'$ is close to $\Omega=\Omega^{(0)}=0$ such that $\Omega'\ll g$. Then, it follows from equation~(\ref{pertb1}) that
\begin{equation}\label{pertb2}
\Delta_{B} = \frac{\Omega'^3}{g_{\rm EP3}^2}=\frac{\Delta_{\omega}^3}{g_{\rm EP3}^2},
\end{equation}
i.e.,
\begin{equation}\label{pertb3}
\Delta_{\omega} = g_{\rm EP3}^{2/3}{\Delta_{B}}^{1/3},
\end{equation}
where the ${\Delta_{B}}^{1/3}$ dependence is characteristic of the EP3.

From equation~(\ref{pertb3}), we obtain the sensitivity factor at the EP3:
\begin{equation}\label{GEP3}
G_{\rm EP3}\equiv\left(\frac{\Delta_{\omega}}{\Delta_{B}}\right)_{\rm EP3}=\frac{g_{\rm EP3}^{2/3}}{{\Delta_{B}}^{2/3}}.
\end{equation}
It tends to infinity when $\Delta_B\rightarrow 0$, indicating the singular behavior at the EP3. In figure~\ref{f3}(d), we numerically show the relation between the sensitivity factor ${G}_{\rm EP3}$ and the perturbation $\Delta_B$.

(ii)~Away from the EP3, there is $g\neq g_{\rm EP3}\equiv 2\gamma/\sqrt{3}$. For a small perturbation, we can still assume that $\Omega'\ll g$ and then equation~(\ref{pertb}) is reduced to
\begin{equation}
\Delta_{B} = -\frac{\left( 3g^{2}-4\gamma^{2} \right)\Omega'}{g^{2} - \left( 3g^{2}-4\gamma^{2} \right)},
\end{equation}
which gives a {\it linear} dependence of the eigenfrequency shift on the perturbation $\Delta_{B}$,
\begin{equation}\label{pertb4}
\Delta_{\omega} =\left(1-\frac{g^2}{3g^2-4\gamma^2} \right)\Delta_{B}.
\end{equation}

Figure~\ref{f3}(c) shows the frequency shift $\Delta_\omega$ of the central eigenvalue versus $|\Delta_B|$. Theoretically, at the EP3, this eigenfrequency shift is $\Delta_\omega = g^{2/3}\Delta_B^{1/3}$, while $\Delta_\omega\sim\Delta_B$ away from the EP3. These theoretical results fit well with the experimental data. The sensitivity factor at the EP3 [figure~\ref{f3}(d)] can be defined by
	\begin{equation}\label{Ham6}
    G_{\rm EP3}\equiv\left(\frac{\Delta_\omega}{\Delta_B}\right)_{\rm EP3}=g^{2/3}\Delta_B^{-2/3}.
	\end{equation}
Compared with the case away from the EP3, the eigenvalue shift $\Delta_\omega$ shows a {\it steep} slope at the EP3, yielding a more profound amplification of the eigenfrequency shift under the same perturbation $\Delta_B$.

\subsection{Enhanced sensing using CPA}

In our experiment, the parameters of the system are chosen to achieve the pseudo-Hermitian Hamiltonian, i.e., the Hamiltonian (\ref{Ham2}) with $g=\sqrt{\Delta^2+\gamma^2}$, but the CPA condition is not obeyed in some cases for comparison. With these parameters, it follows from equation~(\ref{ssum}) in \ref{app3} that the total output spectrum is given by
\begin{align}\label{stot}
&\left| S_{\rm tot}(\Omega) \right|^{2} = \frac{\left( {p + 1} \right)\left| {\left[ {m(\Omega) + {{2\sqrt{\kappa_{1}\kappa_{2}}e^{i\phi}}/\sqrt{p}} + 2\kappa_{1}} \right] + i n(\Omega)} \right|^{2}}{{m(\Omega)}^{2} + {n(\Omega)}^{2}}\nonumber\\
&+ \frac{\left( {\sqrt{p}e^{- i\phi} - \sqrt{\kappa_{1}/\kappa_{2}}} \right)k(\Omega)}{{m(\Omega)}^{2} + {n(\Omega)}^{2}},
\end{align}
with
\begin{align}\label{kPH}
&k(\Omega) = 4\left( {\sqrt{\kappa_{1}\kappa_{2}} + \kappa_{2}e^{i\phi}/\sqrt{p}} \right)^{2}\left( {\sqrt{p}e^{- i\phi} - \sqrt{\kappa_{1}/\kappa_{2}}} \right)\nonumber\\
&+ 4\left( \sqrt{\kappa_{1}\kappa_{2}} + \kappa_{2}e^{i\phi}/\sqrt{p} \right)\left[ {m(\Omega) + 2\sqrt{\kappa_{1}\kappa_{2}}e^{i\phi}/\sqrt{p} + 2\kappa_{1}} \right], \\
&m \left( \Omega \right) = 2\gamma - 2\kappa_{1} - 2\kappa_{2} - \frac{g^{2}\gamma}{\left( {\Omega - \Delta} \right)^{2} + \gamma^{2}} - \frac{g^{2}\gamma}{\left( {\Omega + \Delta} \right)^{2} + \gamma^{2}},\\
&n \left( \Omega \right) = \Omega - \frac{g^{2}\left( {\Omega - \Delta} \right)}{\left( {\Omega - \Delta} \right)^{2} + \gamma^{2}} - \frac{g^{2}\left( {\Omega + \Delta} \right)}{\left( {\Omega' + \Delta} \right)^{2} + \gamma^{2}}, 
\end{align}
where $g=\sqrt{\Delta^2+\gamma^2}$, and $\Omega=\omega-\omega_c$, while $p$ and $\phi$ denote the power ratio and phase difference between the two input field, respectively. When the perturbation is applied, the output spectrum $\vert S'_{\rm tot}\left( \Omega' \right) \vert^{2}$ has the same form as equation~(\ref{stot}), but $\Omega$ is replaced by $\Omega'=\omega'-\omega_c$ and $\pm\Delta$ in equation.~(\ref{kPH}) are replaced by $\Delta+\Delta_B$ and $-\Delta+\Delta_B$, respectively.

In the presence of CPA, it follows from equation~(\ref{a1-a}) that $\sqrt{p}e^{- i\phi}=\sqrt{\kappa_{1}/\kappa_{2}}$. The output spectrum in equation~(\ref{stot}) is reduced to
\begin{equation}\label{speCPA}
\vert {{\rm S}_{\rm tot}\left( \Omega \right)} \vert_{\rm CPA}^{2} = {{\left( {{\kappa_{1}/\kappa_{2}} + 1} \right)\frac{{\Omega}^{2}\left( {\Omega}^{2} - 3g^{2} + 4\gamma^{2} \right)^{2}}{\left( {\Omega}^{2} - g^{2} \right)^{2} + 4{\Omega}^{2}\gamma^{2}}} / \left\lbrack {m\left( \Omega \right)}^{2} + {n\left( {\Omega} \right)}^{2} \right\rbrack}.
\end{equation}
Here $\vert S_{\rm tot}\left( \Omega \right) \vert_{\rm CPA}^{2}$ reaches its minimum value 0 when $\Omega=0$ or $\Omega=\pm\sqrt{3g^2-4\gamma^2}$, which are exactly the eigenvalues of the pseudo-Hermitian Hamiltonian given in equation~(\ref{w1}).

Without the perturbation $\Delta_B$, our system has been tuned to exhibit the CPA, and it can exhibit nearly CPA under a small $\Delta_B$. In decibels, the output spectrum of the system with CPA has an extremely sharp dip, but a less sharp dip occurs in the system nearly with CPA [cf. figure~\ref{f3}(e)].In the present experiment, we focus on the central branch of the three eigenvalues, i.e., $\Omega=\Omega^{(0)}=0$. We can define a sensitivity factor related to the CPA:
\begin{equation}\label{GCPA1}
G_{\rm{CPA}} \equiv 2\pi\frac{\Delta|S_{\rm tot}^{({\rm min})}|^2}{\Delta_{\omega}}= \frac{10{\rm log_{10}}|{S'}_{\rm tot}^{({\rm min})}(\Delta_{\omega})|^2- 10{\rm log_{10}}|S_{\rm tot}^{({\rm min})}(0)|^2_{\rm CPA}}{(\Delta_{\omega}/2\pi)},
\end{equation}
where $|S_{\rm tot}^{({\rm min})}(0)|^2_{\rm CPA}$ is the minimum value of the output spectrum in the presence of CPA, where $\Omega=0$, and $|{S'}_{\rm tot}^{({\rm min})}(\Delta_{\omega})|^2$ is the minimum value of the corresponding output spectrum under the perturbation $\Delta_B$, where $\Omega'=\Delta_{\omega}$.
Because $|{S'}_{\rm tot}^{({\rm min})}(\Delta_{\omega})|^2\neq 0$ and $|S_{\rm tot}^{({\rm min})}(0)|^2_{\rm CPA}=0$, $10\log_{10}|S_{\rm tot}^{({\rm min})}(0)|^2_{\rm CPA}\rightarrow -\infty$, giving rise to $G_{\rm{CPA}}\rightarrow\infty$ in the ideal case. However, in a practical experiment, $10\log_{10}|S_{\rm tot}^{({\rm min})}(0)|^2_{\rm{CPA}}$ cannot reach $-\infty$, but it can reach $-91.5$~dB here at the EP3, resulting from the limited measurement precision of the VNA and the limited adjusting accuracy of the system parameters. Nevertheless, this minimum value can still produce a significantly large enhancement factor $G_{\rm CPA}$ at the EP3 [cf. figure~\ref{f3}(f)], especially in the case of small $\Delta_\omega$.

\subsection{Synthetic sensitivity in the presence of both CPA and EP3}

\begin{figure}
	\centering
	\includegraphics[width=0.95\textwidth]{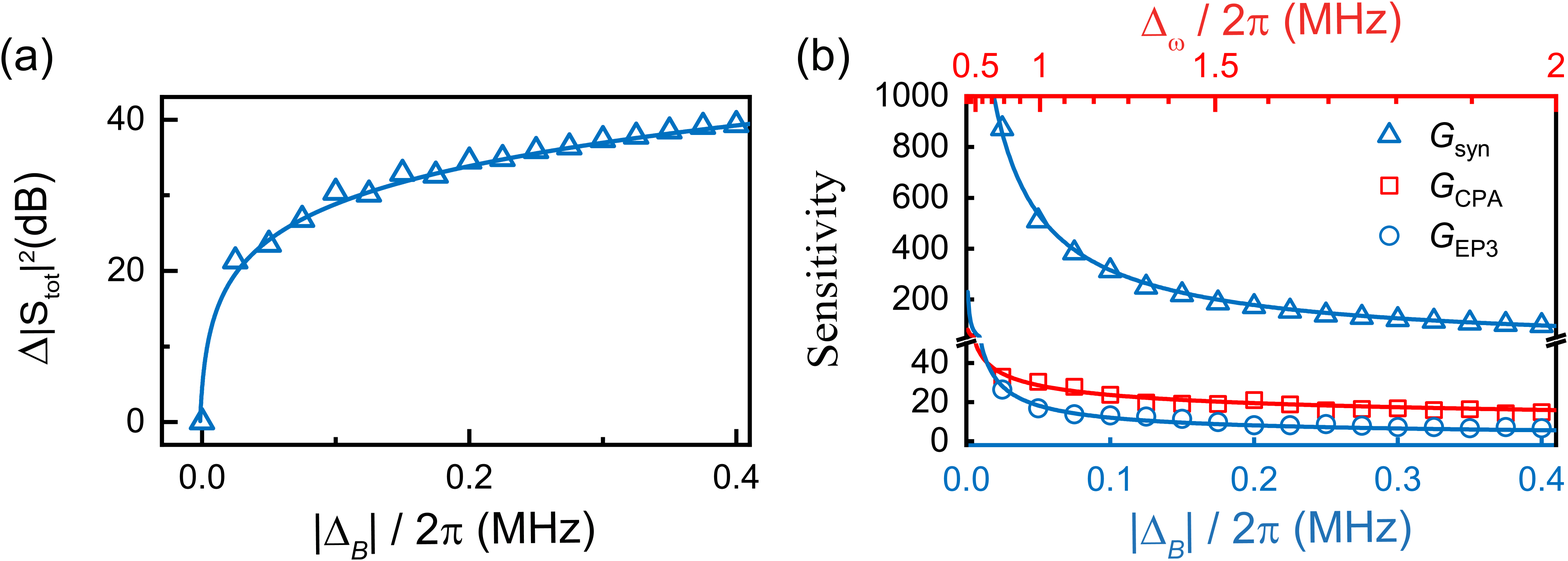}
	\caption{Synthetically enhanced sensitivity in the presence of both CPA and EP3. (a) Minimum-value changes of the output spectra $\Delta|S_{\rm tot}^{({\rm min})}|^2$ at the EP3 versus the perturbation $\Delta_B/2\pi$. (b) Synthetic sensitivity factor $G_{\rm syn}$ in comparison with the respective sensitivity factors $G_{\rm CPA}$ and $G_{\rm EP3}$. Curves are the corresponding fitting results.}\label{f4}
\end{figure}

Based on the experimental data of the output spectra, we obtain the observed changes of the minimum values of the output spectra $\Delta|S_{\rm tot}^{({\rm min})}|^2$ at the EP3 versus the perturbation $\Delta_B$, as shown in figure~\ref{f4}(a). From this relationship between $\Delta|S_{\rm tot}^{({\rm min})}|^2$ and $\Delta_B$, we can define the synthetically enhanced sensitivity of the system in the presence of both CPA and EP3, $G_{\rm syn}\equiv2\pi\Delta|S_{\rm tot}^{({\rm min})}|^2/\Delta_B$, which is readily given by $G_{\rm syn}=G_{\rm CPA}G_{\rm EP3}$. Figure~~\ref{f4}(b) shows the obtained $G_{\rm syn}$ versus the perturbation $\Delta_B$, in comparison with the corresponding $G_{\rm CPA}$ and $G_{\rm EP3}$. Because the enhancements related to the CPA and EP3 are combined in a {\it product} form, the synthetic enhancement factor becomes significantly greater than the respective enhancement factors [figure~\ref{f4}(b)]. This clearly demonstrates the distinct advantage of using both CPA and EP3 in a single non-Hermitian system to produce a sensor with high sensitivity.

\subsection{Detectable minimum change of the magnetic field}
Finally, we discuss the limit of the minimum magnetic-field change that can be detected by the current synthetically enhanced sensor. For the YIG sphere, the relation between the applied magnetic field and the frequency of the uniform magnon mode is
\begin{equation}
\omega_m=\gamma_e|\vec{B}|+\omega_{m,0},
\end{equation}
where $\gamma_e/2\pi=28$~GHz/T denotes the gyromagnetic ratio and $\omega_{m,0}$ is determined by the anisotropy field.
In the synthetically enhanced sensor, the adjustment of the magnetic field is limited by the smallest change of the current supply, i.e., $1\times 10^{-4}$~A, in our electromagnet. The measured frequency change of the magnon mode is $\Delta_B/2\pi=(\omega'_m-\omega_m)/2\pi\approx0.025~{\rm MHz}$, where $\omega'_m$ ($\omega_m$) is the frequency of the magnon mode with (without) the perturbation. It follows from $\Delta_B=\gamma_e\delta{B}$ that the corresponding 
magnetic-field change is $\delta{B}\approx 9\times10^{-7}$~T.

According to equation~(\ref{pertb3}), the frequency shift of the eigenvalue at EP3 is
\begin{equation}\label{w0.6}
\Delta_{\omega} = {g_{\rm EP3}}^{2/3}{\Delta_{B}}^{1/3} = \left( 2\gamma/\sqrt{3} \right)^{2/3} \times (2\pi\times 0.025)^{1/3} \approx 2\pi\times 0.67~({\rm MHz}),
\end{equation}
where $\gamma/2\pi\approx 3$~MHz is the measured damping rates of the two nearly identical YIG spheres.

In our experiment, the minimum value of the total output spectrum changes 21.5~dB around the EP3 when, e.g., $\Delta_{B}/2\pi=0.025$~MHz, as illustrated in figure~\ref{f3}(e). The synthetic sensitivity factor is
$G_{\rm syn}=21.5/0.025=860$. Correspondingly, the CPA-related sensitivity factor and the sensitivity factor at the EP3 are $G_{\rm CPA}=32.1$ and $G_{\rm EP3}=26.8$, respectively. Indeed, as shown in figure~\ref{f4}(b), the synthetic sensitivity factor is much larger than the respective sensitivity factors. Note that the smallest spectrum change that can be resolved by the vector network analyzer used in our experiment (KEYSIGHT PNA-L Network Analyzer N5232B) is $\delta A\sim 1\times 10^{- 13}$~dB. From the definition of the synthetic sensitivity factor, it follows that the detectable minimum change of the magnetic field $\delta B_{\rm min}$ is determined by
\begin{equation}
G_{\rm syn}=2\pi\frac{\delta A}{\gamma_e\delta B_{\rm min}}.
\end{equation}
Thus, we have
\begin{equation}
\delta B_{\rm min}= \frac{2\pi\times 10^{- 13}}{(2\pi\times28\times 10^3)\times 860}\approx 4.2 \times 10^{- 21}~({\rm T}),
\end{equation}
which indicates that the detectable minimum change of the magnetic-field can reach the magnitude $\sim 10^{- 21}$~T.

\section{Conclusions}

In summary, we have demonstrated a synthetically enhanced magnetic sensor using both CPA and EP3 in a pseudo-Hermitian cavity magnonic system. The sensitivity enhancements associated with these two mechanisms are superimposed in a product manner, yielding a greatly amplified synthetic sensitivity. This synthetically enhanced sensitivity offers a new way to explore promising applications of precision metrology via hybrid non-Hermitian quantum systems.

\ack
	This work is supported by the National Key Research and Development Program of China (No.~2022YFA1405200), National Natural Science Foundation of China (No.~$92265202$, No.~$11934010$, and No.~$12174329$), and the Fundamental Research Funds for the Central Universities (No.~$2021{\rm FZZX}001$-$02$).

\appendix
\setcounter{section}{0} 
  
\section{Dissipation rates of the cavity ports}\label{app1}
The dissipation rates of the cavity ports are tuned by changing the lengths of antenna pins inserted into the cavity. The shorter the antenna pin is inserted into the cavity, the smaller the dissipation rate of the corresponding port is.

\begin{figure}[!htb]
	\centering
	\includegraphics[width=0.6\textwidth]{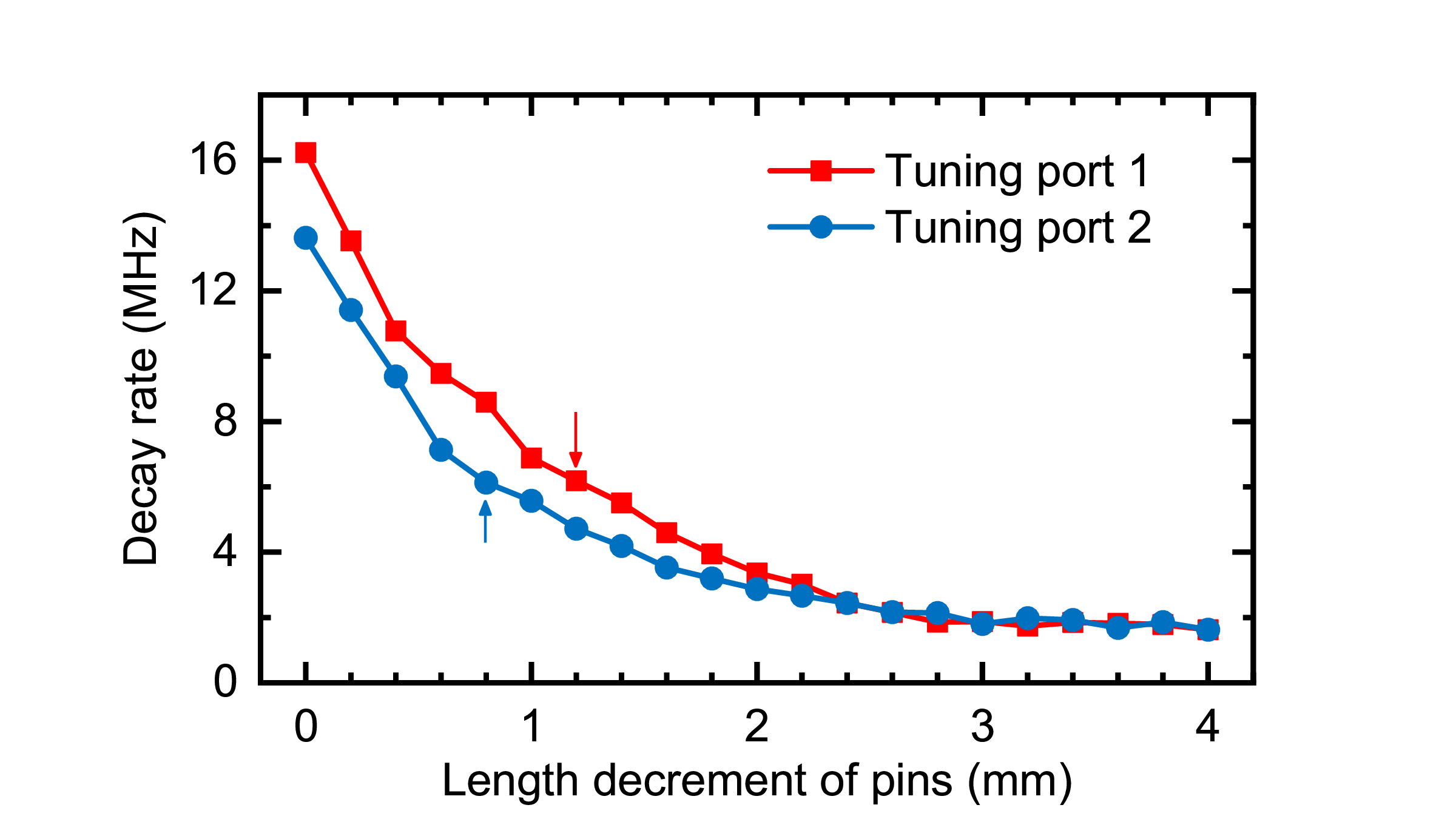}\label{f5}
\caption{Precise adjustments of the dissipation rates of the two ports, i.e., $\kappa_1$ and $\kappa_2$, by tuning the length decrements of the antenna pins. The shorter the pins are inside the cavity, the dissipation rates will be smaller.
When the length decrements are larger than 3~mm, the dissipation rates of the two ports become nearly unchanged. This corresponds to the intrinsic dissipation rate $\kappa_{\rm int}$ of the cavity. The length decrements of the two pins used in our experiment are marked by the red and blue arrows.}
\end{figure}
In figure A1, we gradually reduce the length of the antenna pin inserted into the cavity, and finally the cavity loss converges to a finite value that corresponds to the intrinsic dissipation rate of the cavity, which is about 2~MHz in our experiment. Besides, the dissipation rates of the two ports satisfy $(\kappa_1+\kappa_{\rm int})/2\pi=6$~MHz and $(\kappa_2+\kappa_{\rm int})/2\pi=6$~MHz, as is illustrated with arrows in figure A1.

\section{Non-Hermitian Hamiltonian of the system in the presence of coherent perfect absorption}\label{app2}

In the rotating frame associated with the cavity frequency $\omega_c$, the Hamiltonian can be written as
\begin{equation}\label{H}
H_s/\hbar = \Delta_1b_1^{\dagger}b_1 + \Delta_2b_2^{\dagger}b_2
+ g_1\left( a^{\dagger}b_1 + ab_1^{\dagger} \right) + g_2\left( a^{\dagger}b_2 + ab_2^{\dagger} \right),
\end{equation}
where $\Delta_{1(2)}=\omega_{1(2)}-\omega_c$ is the frequency detuning between the Kittel mode in the YIG sphere 1(2) and the cavity mode. The Langevin equations of the coupled system can now be written as
\begin{align}\label{Lange111}
\dot{a} &= - \frac{i}{\hbar}\left[a,H_s\right] - \left( \kappa_1 + \kappa_2 + \kappa_{\rm int} \right)a + \sqrt{2\kappa_1}a_1^{({\rm in})} + \sqrt{2\kappa_2}a_2^{({\rm in})},\nonumber\\
\dot{b}_1\! &= - \frac{i}{\hbar}\left[b_1,H_s \right] - \gamma_1b_1, \\
\dot{b}_2\! &= - \dfrac{i}{\hbar}\left[b_2,H_s\right] - \gamma_2b_2, \nonumber
\end{align}
where $\kappa_1$, $\kappa_2$ and $\kappa_{\rm int}$ are the dissipation rates of the cavity mode associated with port 1, port 2 and the intrinsic loss of the cavity, respectively, while $\gamma_1$ and $\gamma_2$ are the dissipation rates of the Kittel modes in the two YIG spheres. The input and output fields of the $i$th port, $a_i^{({\rm in})}$ and $a_i^{(\rm out)}$, obey the input-output relation
\begin{equation}
a_i^{({\rm in})} + a_i^{({\rm out})} = \sqrt{2\kappa_i} a,~~~~i = 1,2.
\end{equation}
When the coherent perfect absorption (CPA) occurs, $a_1^{({\rm out})} = a_2^{({\rm out})} = 0$, so
\begin{equation}\label{a1-a}
a_1^{({\rm in})} = \sqrt{2\kappa_1}a,~~~~
a_2^{({\rm in})} = \sqrt{2\kappa_2}a.
\end{equation}
The quantum Langevin equations in equation~(\ref{Lange111}) become
\begin{align}\label{Lange2}
\dot{a} &= - \frac{i}{\hbar}\left[a,H_s\right] + \left( \kappa_1 + \kappa_2 - \kappa_{\rm int} \right)a,\nonumber\\
\dot{b}_1\! &= - \frac{i}{\hbar}\left[b_1,H_s \right] - \gamma_1b_1,\\
\dot{b}_2\! &= - \frac{i}{\hbar}\left[b_2,H_s\right] - \gamma_2b_2, \nonumber
\end{align}
which can be rewritten as
\begin{align}\label{Lange3}
\dot{a} &= - \frac{i}{\hbar}\left[a,H_{\rm eff}\right],\nonumber\\
\dot{b}_1\! &= - \frac{i}{\hbar}\left[b_1,H_{\rm eff} \right],\\
\dot{b}_2\! &= - \frac{i}{\hbar}\left[b_2,H_{\rm eff}\right], \nonumber
\end{align}
where
\begin{equation}
H_{\rm eff}/\hbar = i\kappa_ca_1^{\dagger}a_1+(\Delta_1-i\gamma_1)b_1^{\dagger}b_1 + (\Delta_2-i\gamma_2)b_2^{\dagger}b_2
+ g_1\left( a^{\dagger}b_1 + ab_1^{\dagger} \right) + g_2\left( a^{\dagger}b_2 + ab_2^{\dagger} \right),
\end{equation}
with $\kappa_c\equiv\kappa_1 + \kappa_2 - \kappa_{\rm int}$. When $\kappa_c>0$, the CPA provides an effective gain to the cavity.
In the matrix form, this non-Hermitain Hamiltonian can be written as
\begin{equation}
H_{\rm eff}/\hbar = \begin{pmatrix}
{i\kappa_{c}} & g_{1} & g_{2} \\
g_{1} & {\Delta_{1} - i\gamma_{1}} & 0 \\
g_{2} & 0 & {\Delta_{2} - i\gamma_{2}} \\
\end{pmatrix},
\end{equation}. 

\section{Formulism of the output spectrum}\label{app3}

The quantum Langevin equations of the cavity magnonic system have been illustrated in equation (\ref{Lange111}). It notes that when applying the Fourier transformations
\begin{equation}\label{Frier}
x(t) = \frac{1}{\sqrt{2\pi}}\int_{- \infty}^{+ \infty}\!\!x(\omega)e^{- i\omega t}d\omega,~~~x = a,b_{1},b_{2},
\end{equation}
we obtain the Langevin equations in the frequency domain,
\begin{align}\label{wLange1}
&-i\omega a(\omega) = - \frac{i}{\hbar}\left\lbrack {a(\omega),H_{s}(\omega)} \right\rbrack - \left( {\kappa_{1} + \kappa_{2} + \kappa_{\rm int}} \right)a(\omega)\nonumber\\
& + \sqrt{2\kappa_{1}}a_{1}^{({\rm in})}(\omega) + \sqrt{2\kappa_{2}}a_{2}^{({\rm in})}(\omega), \nonumber\\
&-i\omega b_j(\omega) = - \frac{i}{\hbar}\left\lbrack b_j(\omega),H_s(\omega) \right\rbrack - \gamma_jb_j(\omega),~~j=1,2. 
\end{align}

Using the input-output relation
\begin{equation}\label{IOrela}
a_j^{({\rm in})}(\omega) + a_j^{({\rm out})}(\omega) = \sqrt{2\kappa_j}a(\omega),~~j = 1,2,
\end{equation}
we can obtain the general relations between the output fields, i.e., $a_{1}^{({\rm out})}$ and $a_{2}^{({\rm out})}$, and the input fields, i.e., $a_{1}^{({\rm in})}$ and $a_{2}^{({\rm in})}$:
\begin{align}\label{a1}
&a_{1}^{({\rm out})} = r_{11} a_{1}^{({\rm in})} + t_{12} a_{2}^{({\rm in})},\nonumber\\
&a_{2}^{({\rm out})} = r_{22} a_{2}^{({\rm in})} + t_{21} a_{1}^{({\rm in})},
\end{align}
with the transmission and reflection coefficients
\begin{align}\label{t21}
&t_{12}(\omega)=t_{21}(\omega) = - \frac{2\sqrt{\kappa_{1}\kappa_{2}}}{m + in}, \nonumber\\
&r_{jj}(\omega) = - 1 - \frac{2\kappa_j}{m + in},~~j=1,2,
\end{align}
where
\begin{align}\label{m}
&m = - \left( {\kappa_{1} + \kappa_{2} + \kappa_{\rm int}} \right) - \sum_{j=1}^2\frac{g_j^{2}\gamma_j}{\left( {\omega - \omega_j} \right)^2 + \gamma_j^2}, \nonumber\\
&n = \left( {\omega - \omega_{c}} \right) - \sum_{j=1}^2\frac{g_j^2\left( {\omega - \omega_j} \right)}{\left( {\omega - \omega_j} \right)^2 + \gamma_j^2}.
\end{align}
Note that the output field of each port is the sum of the reflection signal at that port and the transmission signal from the other port. The amplitudes of the output fields at the two ports of the cavity are
\begin{align}\label{s1}
&{\rm S}_{1}(\omega) = r_{11}(\omega) \sqrt{p}e^{- i\phi} + t_{12}(\omega), \nonumber\\
&{\rm S}_{2}(\omega) = r_{22}(\omega) + t_{21}(\omega) \sqrt{p}e^{- i\phi},
\end{align}
where $p$ and $\phi$ denote, respectively, the power ratio and phase difference between the two input fields.

The total output spectrum is obtained by summing the output-field intensities at the two ports:
\begin{align}\label{ssum}
\left| {\rm S_{\rm tot}(\omega)} \right|^{2} &\equiv \left| {\rm S_{1}(\omega)} \right|^{2} + \left| {\rm S_{2}(\omega)} \right|^{2}\notag\\
&= \frac{\left( {p + 1} \right)\left| {\left[ {m(\omega) + {{2\sqrt{\kappa_{1}\kappa_{2}}e^{i\phi}}/\sqrt{p}} + 2\kappa_{1}} \right] + in(\omega)} \right|^{2}}{{m(\omega)}^{2} + {n(\omega)}^{2}}\notag\\
&+\frac{\left( {\sqrt{p}e^{- i\phi} - \sqrt{\kappa_{1}/\kappa_{2}}} \right)k(\omega)}{{m(\omega)}^{2} + {n(\omega)}^{2}},
\end{align}
where
\begin{align}\label{k}
k(\omega) &= 4\left({\sqrt{\kappa_{1}\kappa_{2}} + \kappa_{2}e^{i\phi}/\sqrt{p}} \right)^{2}\left( {\sqrt{p}e^{- i\phi} - \sqrt{\kappa_{1}/\kappa_{2}}} \right) \notag\\&+ 4\left( \sqrt{\kappa_{1}\kappa_{2}} + \kappa_{2}e^{i\phi}/\sqrt{p} \right)\left( {m + 2\sqrt{\kappa_{1}\kappa_{2}}e^{i\phi}/\sqrt{p} + 2\kappa_{1}} \right).
\end{align}.

\end{document}